\documentclass{jfm}
\usepackage{graphicx}
\usepackage{epstopdf, epsfig}
\usepackage{ar}
\usepackage{amsmath,color,graphicx,amssymb}
\usepackage[english]{babel}
\usepackage{color}

\definecolor{purple(munsell)}{rgb}{0.62, 0.0, 0.77}

\shorttitle{Scaling Laws for Heaving and Pitching Propulsors} 
\shortauthor{F. Ayancik, A. Mivehchi, \& K. W. Moored} 

\title{Scaling Laws for Three-Dimensional Combined Heaving and Pitching Propulsors}

\author
 {
 Fatma Ayancik\aff{1}
  \corresp{\email{faa214@lehigh.edu}}, Amin Mivehchi\aff{1}, and Keith W. Moored\aff{1}
  }

\affiliation
{
\aff{1}
Department of Mechanical Engineering, Lehigh University, Bethlehem, PA 18015, USA
}

\begin{document}

\maketitle

\begin{abstract}
We present new scaling laws for the thrust production and power consumption of three-dimensional combined heaving and pitching hydrofoils by extending the three-dimensional pitching scaling laws introduced by \cite{ayancik2019scaling}.  New self-propelled inviscid simulations and previously published experimental data are used to validate the scaling laws over a wide range of motion amplitudes, Strouhal numbers, heave ratios, aspect ratios, and pitching axis locations.  The developed scaling laws are shown to predict inviscid numerical data and experimental data well, within $\pm$25\% and $\pm$16\% of the thrust and power data, respectively. The scaling laws reveal that both the circulatory and added mass forces are important when considering a wide range of motion amplitudes and that nonlinear corrections to classic linear theory are essential to modeling the performance across this wide amplitude range. By using the scaling laws as a guide, it is determined that peak efficiencies occur when $A^* > 1$ and for these large amplitude motions there is an optimal $h^*$ that maximizes the efficiency in the narrow range of $0.75 < h^* < 0.94$.  Finally, the scaling laws show that to further improve efficiency in this high-efficiency regime, the aspect ratio and dimensionless amplitude should be increased, while the Lighthill number should be decreased (lower drag and/or a larger propulsor planform area to wetted surface area ratio), and the pitch axis should be located behind the leading edge.  This scaling model can be used to guide the design of the next generation of high-efficiency bio-inspired machines.
\end{abstract}

\begin{keywords}

\end{keywords}

\section{Introduction}
The engineering of fast, efficient, maneuverable, and quiet bio-inspired propulsive systems has spurred scientific interest in recent years into investigating the unsteady hydrodynamics of fish swimming. Researchers have detailed the complex flow features that are correlated with efficient thrust production \citep{buchholz2008wake, borazjani2008numerical,borazjani2009numerical, masoud2010resonance,dewey2012relationship,moored2014linear,moored2012hydrodynamic,mackowski2015direct,king2018,brooks2019}, revealing that an essential pursuit is a deeper understanding of the origins of unsteady hydrodynamic forces. In this context, some researchers have distilled these flow phenomena into scaling laws under fixed-velocity, net-thrust conditions \citep{green2011unsteady,kang2011effects,dewey2013scaling,quinn2014unsteady,quinn2014scaling,das2016existence}, or for self-propelled organisms \citep{ bainbridge1958speed}.

The basis of many recent scaling laws lie in classic unsteady linear theory. The theories of \cite{theodorsen1935general}, \cite{garrick1936propulsion}, and \cite{mccune1993perspective} have become particularly useful in this pursuit due to their clear assumptions (incompressible and inviscid flow, small-amplitude motions, non-deforming and planar wakes) and the identification of the physical origins of their terms.  For instance, these theories decompose the forces acting on unsteady foils into three types: added mass, quasi-steady, and wake-induced forces. Theodorsen's theory was extended by \cite{garrick1936propulsion} by accounting for the singularity in the vorticity distribution at the leading edge to determine the thrust force produced and the power required by such motions. By following \cite{garrick1936propulsion}, \cite{dewey2013scaling} and \cite{quinn2014unsteady,quinn2014scaling} scaled the thrust forces of pitching and heaving flexible panels with their added mass forces. \cite{moored2018inviscid} advanced this previous work by considering the circulatory and added mass forces of self-propelled pitching foils as well as wake-induced nonlinearities that are not accounted for in classical linear theory \cite[]{garrick1936propulsion}. It was shown that data generated from a potential flow solver was in excellent agreement with the proposed scaling laws. Similarly, \cite{floryan2017scaling} considered both the circulatory and added mass forces and showed excellent collapse of experimental data with their scaling laws for the thrust and power of a heaving or pitching two-dimensional rigid foil.  Following that work, \cite{van2018scaling} developed scaling relations for two-dimensional foils undergoing combined heaving and pitching motions.  While these studies have
provided great insights into the origins of unsteady force production, they have been limited to two-dimensional propulsors.

The scaling laws proposed by \cite{moored2018inviscid} were extended to three-dimensional pitching propulsors \cite[]{ayancik2019scaling} of various aspect ratios by accounting for the added mass of a finite-span propulsor, the downwash/upwash effects from the trailing vortex system, and the elliptical topology of shedding trailing-edge vortices. It was demonstrated that the previous two-dimensional scaling laws as well as their three-dimensional enhancement collapsed both potential flow numerical data as well as experimental data.  Later, \cite{ayancik2020cetacean} developed scaling relations for self-propelled three-dimensional cetacean propulsors undergoing \textit{large-amplitude} combined heaving and pitching motions, which were verified through the use of potential flow numerical data.  By using these scaling relations as a guide, it was demonstrated that the added mass forces played an essential role in understanding the variation in the efficiency with aspect ratio, however, circulatory forces play the predominant role in understanding the variation in the thrust and power with aspect ratio.

Here, we advance the scaling relations introduced in \cite{ayancik2019scaling} and \cite{ayancik2020cetacean} to develop new scaling laws for three-dimensional combined heaving and pitching propulsors valid over a \textit{wide range of amplitudes} and verified with both numerical and experimental data.  This paper is organized in the following manner. Section \ref{sec:problem_formulation} describes the problem formulation and variable ranges for the current study. Section \ref{sec:exp_num_methods} describes the numerical methodology, gives details about the boundary element method and briefly describes the methodology of the previously published experimental data used to verify the scaling laws. Section \ref{sec:scaling} outlines the development of the scaling laws.  Section \ref{sec:results} presents the verification of the scaling laws through new potential flow numerical simulations and previously published experimental data.  Finally, Section \ref{sec:scaling_module} analyzes the scaling laws to derive physical insights for bio-inspired locomotion. 


\section{Problem Formulation}\label{sec:problem_formulation}
\subsection{Idealized Swimmer}
\begin{figure}
    \centering
    \includegraphics[width=1\linewidth]{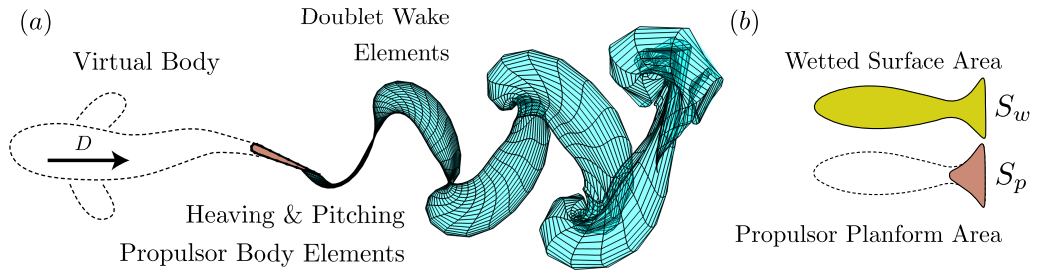}
    \caption{(a) Illustration of an idealized heaving and pitching three-dimensional swimmer as a combination
of a virtual body and propulsor. The doublet wake elements model vorticity shed from the
trailing edge of the propulsor. (b) Representation of the wetted surface area and propulsor
planform area.}
    \label{fig:virtualbody}
\end{figure}
Self-propelled simulations are performed on an idealized swimmer that is a combination of a virtual body and a three-dimensional propulsor (Figure \ref{fig:virtualbody}). The propulsor undergoing combined heaving and pitching motions has a rectangular planform shape,  a NACA 0012 cross-sectional profile, and a chord length of $c = 0.1$ m. The propulsor planform area is $S_p = sc$ where $s$ is the span length of the propulsor. The aspect ratio is then defined as $\AR = s/c$, which varies from 1 to 1000 in the current study, where the highest aspect
ratio represents an effectively two-dimensional propulsor. A drag force, $D$, is applied to the propulsor representing the effect of a virtual body that is not present in the computational domain. The drag force is determined by using a classic high Reynolds number drag law, where drag is
proportional to the square of the swimming speed $U$ as
\begin{equation}
    D = 1/2\, \rho C_D S_w U^2
\end{equation}
where $\rho$ is the density of fluid, $C_D$ is the drag coefficient and $S_w$ represents the total wetted surface area, which is calculated as the product of the propulsor planform area and the ratio of the wetted area to propuslor planform area, $S_{wp}$. The area ratio is chosen to be 5, 10 and 15 in the current study.

\subsection{Input Variables and Propulsor Kinematics}
During self-propelled locomotion, swimmers reach a cruising condition where the time-averaged
thrust and drag are balanced \citep{saadat2017}. The drag coefficient and area ratio both affect how thrust and drag are balanced on a swimmer, and their combination is represented by the Lighthill number, $Li = C_D S_{wp}$. It represents the propulsor loading during self-propelled swimming, which is akin to wing loading in birds and aircraft during flight. When the $Li$ is high there is high propulsor loading and \textit{vice versa}. Low $Li$ swimmers will swim faster than high $Li$ swimmers for a fixed set of kinematics and propulsor geometry. In this study, by keeping the drag coefficient fixed to $C_D = 0.01$, the Lighthill number is varied from 0.05 to 0.15 by changing the area ratio. The non-dimensional mass of the swimmer is defined as the body mass divided by a characteristic added mass of the propulsor, $m^* = m/\rho S_p c$, and is chosen to be 1. 

The propulsor undergoes sinusoidal combined heaving and pitching motions about its pitching axis where the heave and pitch motion are described as
\begin{equation}
    h(t) = h_0 \sin(2 \pi f t)
    \label{EQ:heavemotion}
\end{equation}
\begin{equation}
    \theta(t) = \theta_0 \sin(2 \pi f t + \psi)
    \label{EQ:pitchmotion}
\end{equation}
\noindent where $h_0$ is the heave amplitude, $f$ is the frequency, $t$ is time, $\theta_0$ is the pitching amplitude, and $\psi$ is the phase angle between pitch and heave in radians and is chosen as $\psi = -\pi/2$, which is characteristic of biological locomotion \citep{fish1998biomechanical}. The total peak-to-peak amplitude 
\begin{equation}
   A = A_{TE}(t^*) = 2 \left[h(t^*) + (c/2)(1 - a) \sin(\theta(t^*))\right]
\end{equation}

\noindent depends upon both the heave and pitch motions and is defined when the trailing-edge amplitude reaches its maximum at the time $t^*$.  The dimensionless pitching axis location is $a$ with $a = -1$ and $1$ representing the leading and trailing edge locations, respectively.  The dimensionless amplitude is then $A^* = A/c$.  The proportion of the total amplitude that is derived from heaving is
\begin{equation} \label{eq:hstar}
    h^* = 2h(t^*)/A,
\end{equation}
\noindent and the proportion derived from pitching is then
\begin{equation} \label{eq:tstar}
    \theta^* = 1 - h^*.
\end{equation}
\noindent The heave ratio $h^*$ is implicitly a heave-to-pitch ratio since it represents both the proportions of the heave and pitch to the total amplitude.  For example, when $h^* = 1$ $(\theta^* = 0)$ the motion is purely heaving, when $h^* = 0.5$ $(\theta^* = 0.5)$ the motion is a perfect balance of heaving and pitching amplitudes, and when $h^* = 0$ $(\theta^* = 1)$ the motion is purely pitching. 

All of the input parameters used in the current study are reported in Table \ref{tab:InputParameters} for the numerical simulations and the experiments. For the new simulations, the frequency, amplitude, and aspect ratio ranges are chosen to produce a dataset that covers the Strouhal number, reduced frequency and aspect ratio ranges that are typical of biological and bio-inspired propulsion \citep{sambilay1990interrelationships,saadat2017}.
\begin{table}
 \begin{center}
  \begin{tabular}{lc}
    \textbf{Computational Input Variables/Parameters:}\vspace{0.1in}\\
    Aspect ratio $(\AR)$  & $1 \leq \AR \leq 1000$ \\
    Heave ratio $(h^*)$ & $0.01 \leq h^* < 0.95$\\
    Lighthill number $(Li)$ & $0.05 \leq Li \leq 0.15$ \\
    Frequency $(f)$ [Hz] & $f = 1$ \\
    Reynolds number $(Re)$   & $Re \rightarrow \infty$ \vspace{0.2in}\\

    \textbf{Experimental Input Variables/Parameters:}\vspace{0.1in}\\
    \underline{Tow tank \cite[]{mivehchi2016heaving,perkins2017rolling}:}\vspace{0.05in}\\
    Aspect ratio $(\AR)$  & $\AR = 4.9$ and $\AR = \infty$ \\
    Heave ratio $(h^*)$ & $0.82 < h^* < 1$\\
    Frequency $(f)$ [Hz] & $0.64 < f < 0.87 $ \\
    Reynolds number $(Re)$   & $Re = 21000$ \vspace{0.1in}\\
    
    \underline{Closed-loop water channel \cite[]{ayancik2019scaling}:}\vspace{0.05in}\\
    Aspect ratio $(\AR)$  & $\AR = 1.0, 1.5$ and $\AR = 2$ \\
    Heave ratio $(h^*)$ & $h^* = 0$\\
    Frequency $(f)$ [Hz] & $0.5 \leq f \leq 2$ \\
    Reynolds number $(Re)$   & $Re = 30000$\\
  \end{tabular} \vspace{0.1in}
  \caption{Input variables and parameters used in the present study.}{\label{tab:InputParameters} }
 \end{center}
\end{table}

\subsection{Output Variables}
All of the  output variables are reported as mean quantities that are time-averaged over the last oscillation cycle and are indicated with an overline as ($\overline{\cdot}$). All mean quantities are taken after a swimmer has reached quasi steady-state swimming, defined as the time when the mean net thrust coefficient is $C_{T}^{\text{net}} \leq 10^{-5}$ that is defined as
\begin{equation}
    C_{T}^{\text{net}} = \frac{(\overline{T} - \overline{D})}{1/2 \rho S_p \overline{U}^2},
\end{equation}
with $T$ being the thrust force, calculated by integrating the $-x$ projection of the pressure forces over the surface of the hydrofoil. At this quasi steady-state condition the mean swimming speed is determined, and the reduced frequency and Strouhal number are defined as
\begin{equation}
    k = \frac{fc}{\overline{U}}, \qquad St = \frac{fA}{\overline{U}}.
\end{equation}

\noindent The time-averaged thrust and power coefficients non-dimensionalized by the added mass forces and added mass power from small amplitude theory \citep{garrick1936propulsion} are defined as,
\begin{equation}
\label{eq:GPandGT}
    C_T = \frac{\overline{T}}{\rho S_p f^2 A^2}, \quad C_P = \frac{\overline{P}}{\rho S_p f^2 A^2 \overline{U}}.
\end{equation}
The mean thrust and power may also be non-dimensionalized by the dynamic pressure, 
\begin{equation}
\label{eq:DPandDT}
    C_T = \frac{\overline{T}}{1/2 \rho S_p U^2}, \quad C_P = \frac{\overline{P}}{1/2 \rho S_p U^3}.
\end{equation}
The two normalizations are related by simple transformations: $C_T^{\text{dyn}} = C_T\, (2 St^2)$ and $C_P^{\text{dyn}} = C_P \, (2 St^2)$. 

\section{Numerical and Experimental Methods} \label{sec:exp_num_methods}

\subsection{Unsteady Boundary Element Method}
An unsteady boundary element method (BEM) is employed to model and calculate the forces acting on three-dimensional self-propelled propulsors. The flow is assumed to be incompressible, irrotational and inviscid such that flow is governed by Laplace's equation as $\nabla^2 \phi^* = 0$, where $\phi^*$ is the perturbation potential in a ground-fixed inertial frame of reference. There is a general solution to Laplace's equation subject to (a) a no-flux boundary condition on the surface of the propulsor and (b) a far field boundary condition that flow perturbations must decay with distance from the propulsor. Then, the general solution of Laplace's equation is reduced to finding a distribution of doublet and source elements on the propulsor's surface and wake by satisfying a Dirichlet boundary condition on the surface of the propulsor at each time step. The far-field boundary condition is implicitly satisfied by the elementary solutions of the doublet and source elements. 

To solve the problem numerically, the propulsor and wake surface are discretized by a finite number of quadrilateral boundary elements. Each element on the body surface has an associated collocation point located at the element's center, just inside the body where the Dirichlet condition is enforced. An explicit Kutta condition is enforced at the trailing edge where the vorticity is set to zero. At each time step a wake doublet element is shed with a strength that satisfies Kelvin's circulation theorem. The wake elements are advected with the local velocity field by applying the desingularized Biot-Savart law \citep{krasny1986desingularization} leading to wake deformation and roll up. The tangential perturbation
velocity over the body is found by a local differentiation of the perturbation potential. The unsteady Bernoulli equation is then used to calculate the pressure field acting on the body. Finally, the forces acting on the heaving and pitching propulsor are calculated by an integration of the pressure forces over its boundary. The self-propelled body dynamics is satisfied through a single degree of freedom equation of motion that allows the streamwise translation
of the propulsor. The body position and velocity determined at the $(n+1)^{th}$ time step are calculated by a trapezoidal rule and a forward differencing scheme as suggested by \cite{borazjani2009numerical}, respectively:
\begin{align}
    U_0^{n+1} &= U_0^{n} + \frac{F_x^{n}}{M} \Delta t \\
    x_b^{n+1} &= x_b^{n} + \frac{1}{2} ( U_0^{n+1} + U_0^{n})\Delta t
\end{align}
where $F_x^{n}$ is the net force acting on the foil in the streamwise direction at the $n^{th}$ time step, $x_b$ is the body position of the foil and $\Delta t$ is the time step.
More details and validations of the three-dimensional unsteady boundary element method can be found in \cite{moored2018unsteady}. Further validations and applications of the solver can be found in \cite{quinn2014unsteady,fish2016hydrodynamic,akoz2018unsteady}.

\subsection{Experimental Methods}
Experimental data from three previously published studies \cite[]{mivehchi2016heaving,perkins2017rolling,ayancik2019scaling} are used in the current study to validate the proposed scaling laws.  These previous studies used hydrofoils with NACA 0012 profiles, and prescribed sinusoidal motions described by equations \eqref{EQ:heavemotion} and \eqref{EQ:pitchmotion}. 

The first data set from \cite{mivehchi2016heaving} and \cite{perkins2017rolling} was for relatively high heave ratios from $0.82 \leq h^* \leq 1$.  These studies were conducted in a towing tank and had a hydrofoil with a chord length of $0.07\;\text{m}$ at a fixed chord-based Reynolds number of $2.1\times10^4$.  \cite{perkins2017rolling} reported data from an effectively two-dimensional foil with $\AR=\infty$ and \cite{mivehchi2016heaving} reported data from an hydrofoil with $\AR = 4.9$. The pitching axis of the hydrofoils in both studies is at the 1/3-chord location. In these experiments the heave-to-chord ratio was fixed and equal to $h_0/c =1$, the maximum pitching amplitude varied from $15^\circ$ to $45^\circ$ with an increment of $5^\circ$, and the frequency varied from 0.64 to 0.87 Hz to generate the Strouhal range of $0.3 \leq St \leq 0.5$.

The second data set from \cite{ayancik2019scaling} had a fixed heave ratio of $h^*=0$ throughout the study.  The experiments were conducted in a recirculating water channel. The hydrofoils used in the study had a chord length of $0.1\; \text{m}$ and a fixed chord-based Reynolds number of $Re = 3 \times 10^4$, for three aspect ratios of $\AR = 1.0, 1.5$ and $2.0$.  The pitching frequency was varied from 0.5 to 2.0 Hz in increments of 0.25 Hz  with the pitching axis located at the leading edge.  The non-dimensional peak-to-peak amplitude ($A^*$) varied from 0.2 to 0.5 in intervals of 0.1.

All the relevant input variables and parameters for the experimental studies are given in Table \ref{tab:InputParameters}. More details of their methodologies can be found in \cite{mivehchi2016heaving}, \cite{perkins2017rolling}, and \cite{ayancik2019scaling}.

\section{Extending Garrick's Linear Theory with Nonlinear Corrections}\label{sec:scaling}
\subsection{Garrick's Theory} \label{sec:Garrick}
We begin our scaling analysis with Garrick's full solution \citep{garrick1936propulsion} for the thrust and power coefficients of a combined heaving and pitching propulsor. 
\begin{small}
\begin{align}
    \label{eq:Garrick_thrust_star}
    &C_T = c_1'\,  \underbrace{\frac{4h_0^2}{A^2} [F^2 + G^2]}_{\text{heave}} + c_2' \,  \underbrace{\frac{4 c h_0 \theta_0}{A^2} \left[-(F^2 + G^2)\left(\frac{1}{\pi k}\right) + \frac{G}{2} + \frac{F}{2\pi k}\right]}_{\text{heave + pitch}} \\ \nonumber
    &+ c_3' \, \underbrace{\frac{4 c^2 \theta_0^2}{A^2} \left\{(F^2 + G^2) \left[\frac{1}{\pi^2 k^2} +\left(\frac{1}{2} - a\right)^2\right] + \left(\frac{1}{4} - \frac{a}{2} \right) -\left(\frac{1}{2} - a\right)F -\frac{F}{\pi^2 k^2} - \left(\frac{1}{2} + a\right)\frac{G}{\pi k} \right\}}_{\text{pitch}},
\end{align}
\end{small}
\begin{small}
\begin{align}
    \label{eq:Garrick_power_star}
     &C_P =  c_4' \, \underbrace{\frac{4 h_0^2}{A^2} \left(F\right)}_{\text{heave}} + c_5' \, \underbrace{\frac{4 c h_0 \theta_0}{A^2} \left( \frac{F}{\pi k} - G \right)}_{\text{heave + pitch}}\\\nonumber
     &+ c_6' \, \underbrace{\frac{4 c^2 \theta_0^2}{A^2} \left\{\frac{1}{2}\left(\frac{1}{2} - a\right) - \left(a + \frac{1}{2} \right)\left[F\left(\frac{1}{2} - a\right) + \frac{G}{\pi k} \right] \right\}}_{\text{pitch}}.
\end{align}
\end{small}
\noindent Here, $F$ and $G$ are the real and imaginary parts of Theodorsen's lift deficiency function, respectively \citep{theodorsen1935general}.  The thrust and power are decomposed into their purely pitching, purely heaving and combined heaving and pitching terms as denoted by the underbrackets.  The coefficients have exact values from theory of $c_1'= c_2'= c_4' = c_5' = \pi^3/2$ and $c_3'= c_6' = \pi^3/8$. Garrick's theory makes further assumptions that the motion is of small amplitude, and that the wake is non-deforming and planar.  However, the simulations and experiments in this study vary from small to large amplitude motions and the wakes are deforming.  Therefore, in order to more accurately produce a scaling model relevant to the current data sets, the exact theoretical coefficients are relaxed and left to be determined. Now, equations (\ref{eq:Garrick_thrust_star}) and (\ref{eq:Garrick_power_star}) can be written in a more compact form with $4 h_0^2/A^2 = {h_0^*}^2$, $4 c h_0 \theta_0/A^2 = h_0^*\theta_0^*$, and $4 c^2 \theta_0^2/A^2 = {\theta_0^*}^2$ being substituted,
\begin{eqnarray}
    \label{eq:thrust_approx}
    C_T = c_1\,  {h_0^*}^2 w_1(k) + c_2 \, h_0^*\theta_0^* w_2(k) + c_3 \, {\theta_0^*}^2 w_3(a,k),
\end{eqnarray}
\begin{eqnarray}
    \label{eq:power_approx}
     C_P =  c_4 \, {h_0^*}^2 w_4(k) + c_5 \,  h_0^*\theta_0^* w_5(k) + c_6 \, {\theta_0^*}^2 w_6(a,k).
\end{eqnarray}

The substituted variables approximate the heave ratio and pitch ratio as $h_0^* \approx h^*$ and $\theta_0^* \approx \theta^*$ in the limit of small amplitude motions \citep{floryan2018efficient}, however, the current study examines a wide range of motion amplitudes from small to large, and consequently $h_0^*$ and $\theta_0^*$ must be determined from $h^*$ following equations (\ref{EQ:heavemotion}) - (\ref{eq:tstar}). 

\subsection{Two-Dimensional Scaling Laws}
In order to consider the scaling of performance for combined heaving and pitching two-dimensional propulsors, Garrick's combined heaving and pitching linear theory is extended with nonlinear corrections similar to those proposed by \cite{moored2018inviscid} for pitching two-dimensional propulsors. Garrick's linear theory is modified by three nonlinear corrections: a form drag term, a large-amplitude separating shear layer term and a vortex proximity term, however, the latter two corrections have been modified from those presented in \cite{moored2018inviscid} to account for combined heaving and pitching motions while the form drag term remains solely dependent upon the pitching portion of the motion.  Now, the new scaling law for the thrust coefficient is,
\begin{small}
\begin{align}\label{eq:FullModelThrust}
    C_T = c_1 \, \underbrace{(1/4 - a/2){\theta_0^*}^2}_{\zeta_1}  + c_2 \, \underbrace{{h_0^*}^2 w_1(k)}_{\zeta_2} + c_3 \, \underbrace{h_0^*\theta_0^* w_2(k)}_{\zeta_3} + c_4 \, \underbrace{{\theta_0^*}^2 w_3'(a,k)}_{\zeta_4} + c_5 \, \underbrace{A_p^*}_{\zeta_5}
\end{align} 
\end{small}
\begin{small}
\begin{align*}
    \mbox{where:} \quad  & w_3'(a,k) = \left\{(F^2 + G^2) \left[\frac{1}{\pi^2 k^2} +\left(\frac{1}{2} - a\right)^2\right] -\left(\frac{1}{2} - a\right)F -\frac{F}{\pi^2 k^2} - \left(\frac{1}{2} + a\right)\frac{G}{\pi k} \right\}\\
    & A_p^ * = 2\sin(\theta_0) 
\end{align*}
\end{small}
\noindent Equation (\ref{eq:FullModelThrust}) presents the thrust coefficient that is a combination of the added mass force from pitching ($c_1 \zeta_1$), and the circulatory forces from heave ($c_2 \zeta_2$), combined heave and pitch ($c_3 \zeta_3$), and pitch ($c_4 \zeta_4$), as well as a form drag term ($c_5 \zeta_5$) that comes from the pitch portion of the motion only since it is physically due to the projected frontal area of the propulsor.  Note that the added mass portion of the force from pitching motions has been separated from the circulatory force for pitching motions thus necessitating the use of the prime on $w_3'(a,k)$.

The new scaling law for the power consumption is, 
\begin{align} \label{eq:FullModelPower}
    C_P = \,  &c_6 \,\underbrace{(1/4 - a/2){\theta_0^*}^2}_{\zeta_6} + \, c_7\, \underbrace{{h_0^*}^2 w_4(k)}_{\zeta_7} + c_8 \,  \underbrace{h_0^*\theta_0^* w_5(k)}_{\zeta_8} + c_9 \, \zeta_9 + c_{10} \, \zeta_{10}.
\end{align}
\begin{small}
\begin{align*}
    \mbox{where:} \; \zeta_9 = &\left(\frac{k^*}{k^* + 1}\right)St_p \theta_0\, g_{9}(h_0^*, \theta_0^*, \theta_0, a),
\end{align*}
\end{small}
\begin{small}
\begin{align*}
    \zeta_{10} = &St^2 k^*\,  g_{10}(h_0^*, \theta_0^*, \theta_0, a),
\end{align*}
\end{small}
and, 
\begin{align*}
g_{9}(h_0^*, \theta_0^*, \theta_0, a) &= {h_0^*}^2 + {h_0^*}{\theta_0^*} \left(\frac{3}{4} - a \right) \cos(\theta_0)  + {\theta_0^*}^2 \left[\frac{1}{4} \left(\frac{1}{2} - a \right)(1 - a)\right] \cos^2(\theta_0) \\
 g_{10}(h_0^*, \theta_0^*, \theta_0, a) &= {h_0^*}^3 + {h_0^*}^2{\theta_0^*} \left[\frac{1}{2} \left(\frac{1}{2} - a \right)(1 - a)\right] \cos(\theta_0) +...\\ \nonumber
    &{h_0^*}{\theta_0^*}^2 \left[\frac{1}{4} \left( 2\left(\frac{1}{2} - a \right)(1 - a) + \left(\frac{1}{2} - a \right)^2 \right)\right] \cos^2(\theta_0) +...\\\nonumber &{\theta^*}^3 \left[\frac{1}{8} \left(\frac{1}{2} - a \right)^2(1 - a) \right] \cos^3(\theta_0).
\end{align*}
Here, $St_p$ and $k^*$ represent the Strouhal number based on the pitching amplitude and a modified reduced frequency, respectively. They are defined as $St_p = (fA_p)/U$ and $k^* = {k}/\left(1 + 4\, St^2\right)$. Similar to the thrust coefficient, the power coefficient is a combination of the added-mass power from pitching  ($c_6\, \zeta_6$), the circulatory power from heave ($c_7\, \zeta_7$), the circulatory power from combined heave and pitch ($c_8 \,\zeta_8$), as well as the circulatory nonlinear correction terms modeling the power due to a large-amplitude separating shear layer ($c_9 \,\zeta_9$) and due to the proximity of the trailing-edge vortex ($c_{10} \,\zeta_{10}$). In contrast to the thrust coefficient, in equation (\ref{eq:FullModelPower}), the pitch-only circulatory term was neglected without much penalty to the data collapse (see Section \ref{sec:results}).  The large-amplitude separating shear layer and vortex proximity power corrections introduced in \cite{moored2018inviscid} have also been modified to account for changes that come with heaving motions (see Appendix \ref{appA} for details).

\subsection{Three-Dimensional Scaling Laws}
To account for three-dimensional effects of the propulsors, the two-dimensional scaling laws presented above will be modified.  First, consider the power scaling law.  Previously for pure pitching motions it was identified \cite[]{ayancik2019scaling} that an elliptical vortex ring correction for the separating shear layer term ($c_9\, \zeta_9$), and the vortex proximity term ($c_{10}\, \zeta_{10}$) was necessary. Now, modifying that correction for heaving motions results in new scaling terms $ \zeta_9'$ and $ \zeta_{10}'$ as,
\begin{align*}
    \zeta_9' = \left(\frac{\gamma k^*}{\gamma k^* + 1}\right) St_p \theta_0\, g_{9}(&h_0^*, \theta_0^*, \theta_0, a) \quad \mbox{and} \quad
    \zeta_{10}' = \gamma St^2 k^*\,  g_{10}(h_0^*, \theta_0^*, \theta_0, a) \\
    \mbox{with:}\quad &\gamma = \frac{1}{2} \left[E(m_2) + \frac{E(m_1)}{\AR \sqrt{4kk^*}} \right]
\end{align*}
Here, $m_1$ and $m_2$ are elliptic moduli where $m_1 = \sqrt{1 - 4\AR kk^*}$ and $m_2 = \sqrt{1 + 4\AR kk^*}$ and $E$ is the complete elliptic integral of the second kind \cite[]{ayancik2019scaling}. For more details on the development of these modifications see Appendix B in \cite{ayancik2019scaling}.

Additionally, classic corrections from aerodynamic and hydrodynamic theory will be applied to the circulatory  \cite[]{prandtl1920theory} and added mass  \cite[]{Brennen1982} terms of the two-dimensional core scaling model as follows:
\begin{align*}
    C_T = c_1 \, \zeta_1 \left(\frac{\AR}{\AR + 1}\right) +  \left( c_2\, \zeta_2 + c_3 \, \zeta_3 + c_4 \, \zeta_4 \right) \left(\frac{\AR}{\AR + 2}\right) + c_5 \, \zeta_5,
\end{align*}
\begin{align*}
    C_P = c_6 \, \zeta_1 \left(\frac{\AR}{\AR + 1}\right) +  \left( c_7\, \zeta_7 + c_8 \, \zeta_8 + c_9 \, \zeta_9' + c_{10} \, \zeta_{10}' \right) \left(\frac{\AR}{\AR + 2}\right) .
\end{align*}

\noindent Dividing by the added mass correction, ${\AR}/({\AR + 1}),$ leads to the following, 
\begin{align*}
    C_T \left(\frac{\AR + 1}{\AR}\right)= c_1 \, \zeta_1 +  \left(c_2\, \zeta_2 + c_3 \, \zeta_3 + c_4 \, \zeta_4\right)\left (\frac{\AR + 1}{\AR + 2}\right) + c_5 \, \zeta_5 \left(\frac{\AR + 1}{\AR}\right)
\end{align*}
\begin{align*}
    C_P \left(\frac{\AR + 1}{\AR}\right)= c_6 \, \zeta_1 +  \left(c_7\, \zeta_7 + c_8 \, \zeta_8 + c_9 \, \zeta_9' + c_{10} \, \zeta_{10}'\right) \left(\frac{\AR + 1}{\AR + 2}\right).
\end{align*}
This reveals a more compact form of the three-dimensional scaling relations as,
\begin{equation}\label{eqn:3D_HP_Ct}
    C_T^* = c_1 \, \zeta_1 + c_2\, \zeta_2^* + c_3 \, \zeta_3^* + c_4 \, \zeta_4^* + c_5 \, \zeta_5^* 
\end{equation}
\begin{equation}\label{eqn:3D_HP_Cp}
    C_P^* = c_6 \, \zeta_6 + c_7\, \zeta_7^* + c_8 \, \zeta_8^* + c_9 \, \zeta_9^* + c_{10} \, \zeta_{10}^*. 
\end{equation}
In the thrust coefficient equation $C_T^* = C_T \left[(\AR + 1)/\AR \right]$, and the various $\zeta^*$ are defined as $\zeta_{2,3,4}^* = \zeta_{2,3,4}[(\AR + 1)/(\AR + 2)]$ and $\zeta_{5}^* = \zeta_{5}[(\AR + 1)/\AR]$. In the power coefficient equation $C_P^* = C_P \left[(\AR + 1)/\AR \right]$, and the various $\zeta^*$ are defined as $\zeta_{7,8}^* = \zeta_{7,8}[(\AR + 1)/(\AR + 2)]$ and $\zeta_{9,10}^* = \zeta_{9,10}'[(\AR + 1)/(\AR + 2)]$.

The scaling relations can also be written in terms of the thrust and power coefficients normalized by dynamic pressure as,
\begin{equation}
    C_T^{{\text{dyn}}^*} = (c_1 \, \zeta_1 + c_2\, \zeta_2^* + c_3 \, \zeta_3^* + c_4 \, \zeta_4^* + c_5 \, \zeta_5^*)\,2St^2
\end{equation}
\begin{equation}
    C_P^{{\text{dyn}}^*} = (c_6 \, \zeta_6 + c_7\, \zeta_7^* + c_8 \, \zeta_8^* + c_9 \, \zeta_9^* + c_{10} \, \zeta_{10}^*)\,2St^2. 
\end{equation}

\section{Results and Discussion}\label{sec:results}
\begin{figure}
    \centering
    \includegraphics[width=1\linewidth]{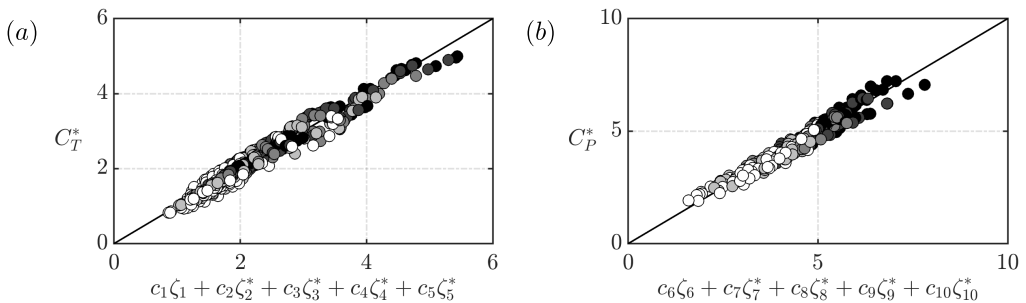}
    \caption{Scaling of the (a) thrust and (b) power coefficient data for all motion amplitudes and aspect ratios considered in the simulations with the thrust equation constants determined to be $c_1 = 2.69$, $c_2 = 18.37$, $c_3 = 3.39$, $c_4 = 19.85$ and $c_5 = -0.16$. The power equation constants were determined to be $c_6 = 5.03$, $c_7 = 19.29$, $c_8 = 10.05$, $c_9 = 11.61$ and $c_{10} = 13.44$.}
    \label{fig:Scaled_Num}
\end{figure}

The computational input parameters (Table \ref{tab:InputParameters}) lead to more than five hundred, three-dimensional, self-propelled simulations with a reduced frequency range of $0.04 < k < 1.32$ and a Strouhal number range of $0.07 < St < 0.31$. From these data points, the thrust and power coefficients as defined in equation (\ref{eq:GPandGT}) are calculated.
\begin{figure}
    \centering
    \includegraphics[width=1\linewidth]{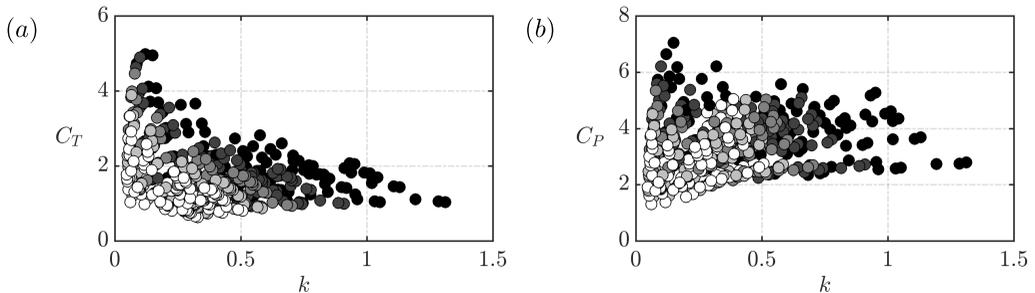}
    \caption{The (a) thrust and (b) power coefficient data as a function of reduced frequency from the self-propelled simulations.}
    \label{fig:Unscaled_Num}
\end{figure}
Figure \ref{fig:Unscaled_Num} presents the
thrust and power coefficients as functions of the reduced frequency for all motion amplitudes, aspect ratios and $Li$ numbers considered here. From black to white the marker colors correspond to pitch angles changing from low to high values, respectively. The thrust and power of a heaving and pitching hydrofoil vary widely with its input parameters.  Figure \ref{fig:Scaled_Num} shows the numerical data plotted as a function of three-dimensional scaling relations proposed in (\ref{eqn:3D_HP_Ct}) and (\ref{eqn:3D_HP_Cp}) and an excellent collapse of the data is observed. 
The collapsed data can be seen to follow a line of slope one for both the thrust and power within $\pm 12\%$ (thrust) and $\pm 10\%$ (power) of the predicted scaling law.  The values of the constants are determined by minimizing the squared residuals between the data and the scaling law prediction.  The constants in the thrust law are $c_1 = 2.69$, $c_2 = 18.37$, $c_3 = 3.39$, $c_4 = 19.85$ and $c_5 = -0.16$, while for the power law they are $c_6 = 5.03$, $c_7 = 19.29$, $c_8 = 10.05$, $c_9 = 11.61$ and $c_{10} = 13.44$.

\begin{figure}
    \centering
    \includegraphics[width=1\linewidth]{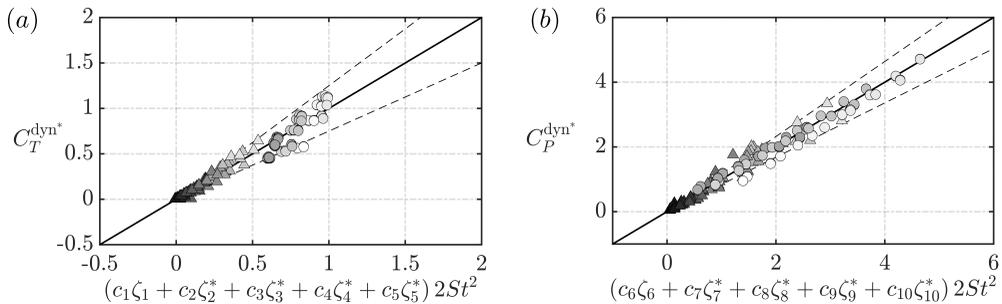}
    \caption{Scaling of the (a) thrust and (b) power coefficient data for all motion amplitudes and aspect ratios considered in the experiments with the thrust equation constants determined to be $c_1 = 3.14$, $c_2 = 7.68$, $c_3 = 8.62$, $c_4 = -20.05$ and $c_5 = -0.79$. The power equation constants were determined to be $c_6 = 7.37$, $c_7 = 24.89$, $c_8 = 18.66$, $c_9 = -369.41$ and $c_{10} = 206.85$.}
    \label{fig:Scaled_Exp}
\end{figure}

To validate that the scaling laws equally apply to viscous flows, the compilation of experimental data is graphed against the scaling law predictions in Figure \ref{fig:Scaled_Exp}.  The experimental data have a reduced frequency range of $0.15 < k < 1.34 $ and a Strouhal number range of $0.06 \leq St < 0.45$. The marker colors represents the pitch angle in the same way as in Figure \ref{fig:Unscaled_Num}. The circle and triangle marker types indicate $Re =$ 20,000 and $Re =$ 30,000, respectively. The dashed lines on the thrust and power graphs represent $\pm 25\%$ and $\pm 16\%$ error, respectively. For both thrust and power, we see a collapse of the data nearly within these limits.  The experimentally determined constants for the thrust equation are $c_1 = 3.14$, $c_2 = 7.68$, $c_3 = 8.62$, $c_4 = -20.05$, $c_5 = -0.79$ and for the power equation they are $c_6 = 7.37$, $c_7 = 24.89$, $c_8 = 18.66$, $c_9 = -369.41$ and $c_{10} = 206.85$.

Our scaling relations show a good collapse of the data for a wide range of Reynolds number from $Re =$ 20,000 and $Re =$ 30,000 in the experiments to $Re = \infty$ in the inviscid simulations. Although, it should be noted that the determined coefficients are different between the simulations and the experiments.  The $c_9$ coefficient switches sign in the experimental data, meaning it represents power extraction, whereas the other coefficients represent power consumption. This is likely due to a viscosity-driven phase shift between the lift term scaled by $c_9$ and the velocity of the hydrofoil that is not captured in the inviscid numerics.  In contrast, the added mass thrust and power terms are positive in both experiments and simulations, as expected based on physical grounds.  To approach universal constants it would be important to account for the Reynolds number variation in the thrust and power data as pioneered by \cite{Senturk2019}.  However, the common agreement in the basic scaling terms for inviscid and viscous flows indicates that the dominant flow physics are inviscid in nature. The small differences between the scaling law agreement in the experiments and the simulations may be attributed to secondary viscous effects.

The collapse of the data to a line of slope one for both numerical and experimental cases confirms that the newly proposed three-dimensional scaling relations capture the dominant flow physics of three-dimensional heaving and pitching propulsors across a wide range of $St$, $A^*$, $h^*$ and $\AR$.

\section{Physical insight derived from
the scaling laws}\label{sec:scaling_module}
In this section, the developed scaling laws are used as a guide to more deeply understand the effects of propulsor aspect ratio and kinematics on the efficiency. To determine a scaling law for efficiency, we simply take the ratio of the thrust and power coefficient scaling laws, which becomes, 
\begin{equation}
    \eta = \frac{ c_1 \, \zeta_1 +  \left(c_2\, \zeta_2 + c_3 \, \zeta_3 + c_4 \, \zeta_4\right)\left[(\AR + 1)/(\AR + 2)\right] + c_5 \, \zeta_5 \left[(\AR + 1)/(\AR)\right]}{c_6 \, \zeta_6 + (c_7\, \zeta_7 + c_8 \, \zeta_8 + c_9 \, \zeta_9' + c_{10} \, \zeta_{10}') \left[(\AR + 1)/(\AR + 2)\right]}.
\end{equation}
The experimentally determined coefficients are used in this section for analyzing the efficiency scaling law (see Section \ref{sec:results} for the coefficient values). 

\begin{figure}
    \centering
    \includegraphics[width=1\linewidth]{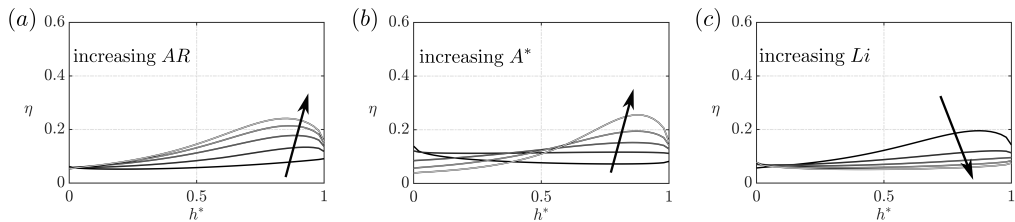}
    \caption{Efficiency as a function of $h^*$ for (a) $A^* = 1$, $1 \leq \AR \leq 5$ and $Li=0.2$, (b) $\AR=3$, $0.2 \leq A^* \leq 1.5$ and $Li=0.2$, (c) $A^*=1$, $\AR=3$ and $0.2\leq Li \leq 1$ when pitching axis at leading edge ($a = -1$). Colours from black to white correspond from small to large values, respectively, for $\AR$, $A^*$ and $Li$.}
    \label{fig:Scaling_M_LE}
\end{figure}

Next, the scaling laws are used to examine the trends in efficiency when the $h^*$, $A^*$, $\AR$, $Li$ number and $a$ are varied independently. As a baseline case, $\AR = 3$, $A^* = 1$ and $Li = 0.2$ are chosen since they correspond to typical values for bio-locomotion and bio-inspired locomotion. Figure \ref{fig:Scaling_M_LE} presents the efficiency as a function of $h^*$ with the pitch axis at the leading edge ($a = -1$).  In Figure \ref{fig:Scaling_M_LE}(a) the lines represent various $\AR$ with $A^* = 1$ and $Li = 0.2$. In Figure \ref{fig:Scaling_M_LE}(b) the lines represent various $A^*$ with $\AR = 3$ and $Li = 0.2$. In Figure \ref{fig:Scaling_M_LE}(c) the lines represent various $Li$ with $A^* = 1$ and $\AR = 3$. Line colors from black to white correspond from small to large values, respectively, for $\AR$, $A^*$ and $Li$.

All figures show that there is an optimal $h^*$ that maximizes the efficiency. Figure \ref{fig:Scaling_M_LE}(a) shows that as the aspect ratio is increased the optimal $h^*$ decreases. In Figure \ref{fig:Scaling_M_LE}(b), we see that to achieve high efficiency, the total amplitude-to-chord ratio $A^*$ should be large. This observation is consistent with the argument put forth by \cite{floryan2018efficient} where they proposed that large-amplitude motions are more efficient than small-amplitude motions. \cite{floryan2018efficient} also stated that optimal efficiency should occur when $h^* = 0.5$ where heaving and pitching motions contribute equally to the total motion. However, Figure \ref{fig:Scaling_M_LE}(b) shows that the optimal value is not near $h^* = 0.5$ in this data, but is, in fact, closer to $h^* = 0.8$ for the most efficient cases. When $A^* < 1$, peak efficiencies occur for pitch-dominated motions ($h^* < 0.5$) and for $A^* \geq 1$ they occur for heave-dominated motions ($h^* > 0.5$). In fact, pitch-dominated motions increase their efficiency with decreasing $A^*$ values and that trend is reversed for heave-dominated motions.  It is also observed that higher $Li$ number swimmers have higher propulsive efficiency for pitch dominated motions and \textit{vice versa} for heave dominated motions. This behavior is a consequence of a chain of effects. As an example, an increase in $Li$ leads to a higher drag producing body, lower swimming speed, higher reduced frequency, and therefore lower efficiency for heavy-dominated motions \citep{garrick1936propulsion, anderson1998oscillating, dewey2013scaling, quinn2014unsteady, moored2018inviscid}.  Lastly, for pitch dominated motions the efficiency is relatively insensitive to changes in $\AR$.  The main conclusion from Figure \ref{fig:Scaling_M_LE} is that the highest efficiencies occur for a narrow range of $h^*$ from $0.75 < h^* < 0.94$ for $A^* \geq 1$.
\begin{figure}
    \centering
    \includegraphics[width=1\linewidth]{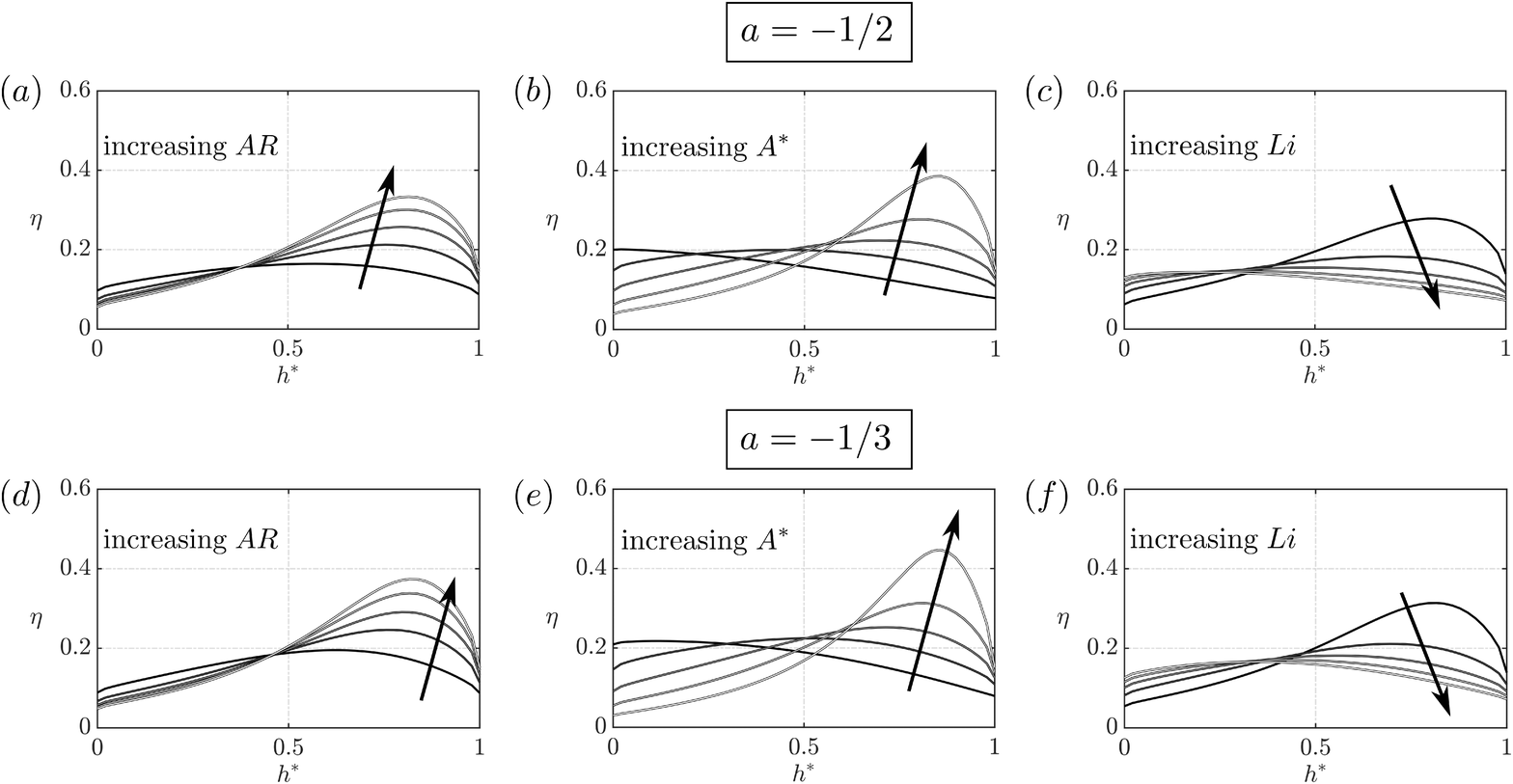}
    \caption{Efficiency as a function of $h^*$ for (a,d) $A^* = 1$, $1 \leq \AR \leq 5$ and $Li=0.2$, (b,e) $\AR=3$, $0.2 \leq A^* \leq 1.5$ and $Li=0.2$, (c,f) $A^*=1$, $\AR=3$ and $0.2\leq Li \leq 1$ when pitching axis at 1/3 chord ($a = -1/3 $) and at mid chord ($a = 0$). Colours corresponds to the same values indicated in figure \ref{fig:Scaling_M_LE}.}
    \label{fig:Scaling_M}
\end{figure}

Figure \ref{fig:Scaling_M} presents the efficiency variation as functions of $h^*$, $\AR$, $A^*$ and $Li$ for different pitch axis locations. Figures \ref{fig:Scaling_M}(a,d), (b,e), and (c,f) indicate the same trends as Figures \ref{fig:Scaling_M_LE}(a), (b), and (c), respectively. Now, its observed that the variation in the pitching axis changes the maximum efficiency peak and optimum $h^*$ value. When the pitch axis moves from the leading edge toward the mid chord, the maximum efficiency increases for all the parameter variations considered here.

\section{Conclusions}
New scaling laws are developed for the thrust generation and power consumption of three-dimensional combined heaving and pitching propulsors by extending the three-dimensional pitching scaling laws introduced by \cite{ayancik2019scaling} to consider combined heaving and pitching motions. The developed scaling laws are shown to predict inviscid numerical data and experimental data well, within $\pm$25\% and $\pm$16\% of the thrust and power data, respectively. The scaling laws reveal that both the circulatory and added mass forces are important when considering a wide range of motion amplitudes and that nonlinear corrections to classic linear theory are essential to modeling the performance across this wide amplitude range. By using the scaling laws as a guide, it is determined that peak efficiencies occur when $A^* > 1$ and for these large amplitude motions there is an optimal $h^*$ that maximizes the efficiency in the narrow range of $0.75 < h^* < 0.94$.  Finally, the scaling laws show that to further improve efficiency in this high-efficiency regime, the aspect ratio and dimensionless amplitude should be increased, while the Lighthill number should be decreased (lower drag and/or a larger propulsor planform area to wetted surface area ratio), and the pitch axis should be located downstream of the leading edge.
\section*{Acknowledgements}
This work was supported by the Office of Naval Research
(ONR) under the Multidisciplinary University Research Initiative (MURI) program (grant no. N00014-14-1-0533).
\appendix
\section*{Declaration of Interest}
The authors report no conflict of interest.

\section{Derivation of Scaling Relations}\label{appA} 
The separating shear layer and vortex proximity terms from \cite{moored2018inviscid} rely on balancing the cross-stream component of the velocity induced at the trailing edge by a shedding trailing-edge vortex and the cross-stream component of the velocity induced by the bound vortex with circulation $\Gamma_b$. This enforces the Kutta condition at the trailing edge and determines a scaling for the additional bound circulation, $\Gamma_b = \Gamma_0 - \Gamma_1$, where $\Gamma_0$ is the bound circulation from the quasi-steady motion of the foil alone, while the additional bound circulation is the bound circulation induced by the influence of the wake. These terms are modified to fully account for combined heaving and pitching motion. We started with the separating shear layer term,
\begin{equation}
    L_{\text{sep}} = \rho s (\Gamma_0 + \Gamma_1) \frac{dx}{dt} \quad \mbox{where:} \, \frac{dx}{dt} = c \dot{\theta} \sin{\theta}
\end{equation}
The dot over some terms corresponds to the time rate of change. When we take the time-average the $\Gamma_0$ term goes to zero due to near-orthogonality between $\dot{\theta}$ and $\sin{\theta}$. Then, 
\begin{equation}\label{eqn:P_sep}
    \overline{P}_{\text{sep}} = \rho s \Gamma_1 c \dot{\theta} \sin{\theta} \,[\dot{h} + {1}/{2}(1-a)c\, \dot{\theta}\cos(\theta)]
\end{equation}
where, $\Gamma_1$ is the additional circulation determined by \cite{moored2018inviscid} as, 
\begin{equation}
    \Gamma_1 \propto - \Gamma_0 \left( \frac{k^*}{k^* + 1} \right).
\end{equation}

For large-amplitude, combined heaving and pitching motions, the bound circulation is defined as,  
\begin{equation}\label{eqn:BC}
   \Gamma_0 \propto - c \left [\dot{h} + c\,\dot{\theta}\cos(\theta)\, \frac{1}{2} \left(\frac{1}{2} - a \right)\right],
\end{equation}
with $a$ corresponding to pitch axis location as defined by \cite{garrick1936propulsion}. 

Now, substituting the relation (\ref{eqn:BC}) into (\ref{eqn:P_sep}), the following power relation for the separating shear layer term can be obtained,
\begin{equation}\label{eqn:P_sep2}
    \overline{P}_{\text{sep}} = \rho S_p c f^3 \theta_0 \sin{\theta_0} \left( \frac{k^*}{k^* + 1} \right) \left[\dot{h} + {1}/{2}(1-a)c\, \dot{\theta}\cos(\theta)\right] \left [\dot{h} + c\,\dot{\theta}\cos(\theta)\, \frac{1}{2} \left(\frac{1}{2} - a \right)\right].
\end{equation}

The second explicitly nonlinear correction to the power is due to
the proximity of the trailing-edge vortex, where the lift from this term is
\begin{equation}
    L_{\text{prox}} = \rho s (\Gamma_0 + \Gamma_1)\, u_{\text{ind}}.
\end{equation}
Since $\Gamma_1$ is a perturbation of $\Gamma_0$, we neglect $\Gamma_1$. Then, 
\begin{equation}
    L_{\text{prox}} = \rho s \Gamma_0  u_{\text{ind}} \quad \mbox{where:} \; u_{\text{ind}} = \frac{\Gamma_0 f St}{U(1 + 4St^2)}
\end{equation}
Consequently, the power scaling from the proximity of the trailing-edge vortex is obtained as
\begin{equation}
    \overline{P}_{\text{prox}} = \frac{\rho S_p c\, f^4 St}{U(1 + 4St^2)}\left[\dot{h} +  {1}/{2}(1-a)c\, \dot{\theta}\cos(\theta)\right] \left [\dot{h} +c\,\dot{\theta}\cos(\theta)\, \frac{1}{2} \left(\frac{1}{2} - a \right)\right]^2.
\end{equation}

\bibliographystyle{jfm}
\bibliography{jfm-instructions}

\begin{thebibliography}{39}
\expandafter\ifx\csname natexlab\endcsname\relax\def\natexlab#1{#1}\fi
\def\au#1{#1} \def\ed#1{#1} \def\yr#1{#1}\def\at#1{#1}\def\jt#1{\textit{#1}}
  \def\bt#1{#1}\def\bvol#1{\textbf{#1}} \def\vol#1{#1} \def\pg#1{#1}
  \def\publ#1{#1}\def\arxiv#1{#1}\def\org#1{#1}\def\st#1{\textit{#1}}

\bibitem[Akoz \& Moored(2018)]{akoz2018unsteady}
{\sc \au{Akoz, E.} \& \au{Moored, K.~W.}} \yr{2018}  \at{Unsteady propulsion by
  an intermittent swimming gait}.  \jt{Journal of Fluid Mechanics}  \bvol{834},
   \pg{149--172}.

\bibitem[Anderson {\em et~al.\/}(1998)Anderson, Streitlien, Barrett \&
  Triantafyllou]{anderson1998oscillating}
{\sc \au{Anderson, J.~M.}, \au{Streitlien, K.}, \au{Barrett, D.~S.} \&
  \au{Triantafyllou, M.~S.}} \yr{1998}  \at{Oscillating foils of high
  propulsive efficiency}.  \jt{Journal of Fluid mechanics}  \bvol{360},
  \pg{41--72}.

\bibitem[Ayancik {\em et~al.\/}(2020)Ayancik, Fish \&
  Moored]{ayancik2020cetacean}
{\sc \au{Ayancik, F.}, \au{Fish, F.~E.} \& \au{Moored, K.~W.}} \yr{2020}
  \at{Three-dimensional scaling laws of cetacean propulsion characterize the
  hydrodynamic interplay of flukes' shape and kinematics}.  \jt{Journal of the
  Royal Society Interface}  \bvol{17}~(163),  \pg{20190655}.

\bibitem[Ayancik {\em et~al.\/}(2019)Ayancik, Zhong, Quinn, Brandes, Bart-Smith
  \& Moored]{ayancik2019scaling}
{\sc \au{Ayancik, F.}, \au{Zhong, Q.}, \au{Quinn, D.~B.}, \au{Brandes, A.},
  \au{Bart-Smith, H.} \& \au{Moored, K.~W.}} \yr{2019}  \at{Scaling laws for
  the propulsive performance of three-dimensional pitching propulsors}.
  \jt{Journal of Fluid Mechanics}  \bvol{871},  \pg{1117--1138}.

\bibitem[Bainbridge(1958)]{bainbridge1958speed}
{\sc \au{Bainbridge, R.}} \yr{1958}  \at{The speed of swimming of fish as
  related to size and to the frequency and amplitude of the tail beat}.
  \jt{Journal of experimental biology}  \bvol{35}~(1),  \pg{109--133}.

\bibitem[Borazjani \& Sotiropoulos(2008)]{borazjani2008numerical}
{\sc \au{Borazjani, I.} \& \au{Sotiropoulos, F.}} \yr{2008}  \at{Numerical
  investigation of the hydrodynamics of carangiform swimming in the
  transitional and inertial flow regimes}.  \jt{Journal of experimental
  biology}  \bvol{211}~(10),  \pg{1541--1558}.

\bibitem[Borazjani \& Sotiropoulos(2009)]{borazjani2009numerical}
{\sc \au{Borazjani, I.} \& \au{Sotiropoulos, F.}} \yr{2009}  \at{Numerical
  investigation of the hydrodynamics of anguilliform swimming in the
  transitional and inertial flow regimes}.  \jt{Journal of Experimental
  Biology}  \bvol{212}~(4),  \pg{576--592}.

\bibitem[Brennen(1982)]{Brennen1982}
{\sc \au{Brennen, C.~E.}} \yr{1982}  \bt{{A review of added mass and fluid
  inertial forces}}. {\em Tech. Rep.\/} January.  \org{Naval Civil Engineering
  Laboratory}, Sierra Madre.

\bibitem[Brooks \& Green(2019)]{brooks2019}
{\sc \au{Brooks, S.~A.} \& \au{Green, M.~A.}} \yr{2019}  \at{Experimental study
  of body-fin interaction and vortex dynamics generated by a two
  degree-of-freedom fish model}.  \jt{Biomimetics}  \bvol{4}~(67),  \pg{1--23}.

\bibitem[Buchholz \& Smits(2008)]{buchholz2008wake}
{\sc \au{Buchholz, J. H.~J.} \& \au{Smits, A.~J.}} \yr{2008}  \at{The wake
  structure and thrust performance of a rigid low-aspect-ratio pitching panel}.
   \jt{Journal of fluid mechanics}  \bvol{603},  \pg{331--365}.

\bibitem[Das {\em et~al.\/}(2016)Das, Shukla \& Govardhan]{das2016existence}
{\sc \au{Das, A.}, \au{Shukla, R.~K.} \& \au{Govardhan, R.~N.}} \yr{2016}
  \at{Existence of a sharp transition in the peak propulsive efficiency of a
  low-$ re $ pitching foil}.  \jt{Journal of Fluid Mechanics}  \bvol{800},
  \pg{307--326}.

\bibitem[Dewey {\em et~al.\/}(2013)Dewey, Boschitsch, Moored, Stone \&
  Smits]{dewey2013scaling}
{\sc \au{Dewey, P.~A.}, \au{Boschitsch, B.~M.}, \au{Moored, K.~W.}, \au{Stone,
  H.~A.} \& \au{Smits, A.~J.}} \yr{2013}  \at{Scaling laws for the thrust
  production of flexible pitching panels.}  \jt{Journal of Fluid Mechanics}
  \bvol{732},  \pg{29--46}.

\bibitem[Dewey {\em et~al.\/}(2012)Dewey, Carriou \&
  Smits]{dewey2012relationship}
{\sc \au{Dewey, P.~A.}, \au{Carriou, A.} \& \au{Smits, A.~J.}} \yr{2012}
  \at{On the relationship between efficiency and wake structure of a
  batoid-inspired oscillating fin}.  \jt{Journal of fluid mechanics}
  \bvol{691},  \pg{245--266}.

\bibitem[Fish(1998)]{fish1998biomechanical}
{\sc \au{Fish, Frank~E}} \yr{1998}  \at{Biomechanical perspective on the origin
  of cetacean flukes}.  \bt{In {\em The emergence of whales\/}},  \pg{pp.
  303--324}.  \publ{Springer}.

\bibitem[Fish {\em et~al.\/}(2016)Fish, Schreiber, Moored, Liu, Dong \&
  Bart-Smith]{fish2016hydrodynamic}
{\sc \au{Fish, F.~E.}, \au{Schreiber, C.~M.}, \au{Moored, K.~W.}, \au{Liu,
  Geng}, \au{Dong, H.} \& \au{Bart-Smith, H.}} \yr{2016}  \at{Hydrodynamic
  performance of aquatic flapping: efficiency of underwater flight in the
  manta}.  \jt{Aerospace}  \bvol{3}~(3),  \pg{20}.

\bibitem[Floryan {\em et~al.\/}(2017)Floryan, Van~Buren, Rowley \&
  Smits]{floryan2017scaling}
{\sc \au{Floryan, D.}, \au{Van~Buren, T.}, \au{Rowley, C.~W.} \& \au{Smits,
  A.~J.}} \yr{2017}  \at{Scaling the propulsive performance of heaving and
  pitching foils}.  \jt{Journal of Fluid Mechanics}  \bvol{822},
  \pg{386--397}.

\bibitem[Floryan {\em et~al.\/}(2018)Floryan, Van~Buren \&
  Smits]{floryan2018efficient}
{\sc \au{Floryan, D.}, \au{Van~Buren, T.} \& \au{Smits, A.~J.}} \yr{2018}
  \at{Efficient cruising for swimming and flying animals is dictated by fluid
  drag.}  \jt{Proceedings of the National Academy of Sciences}
  \bvol{115}~(32),  \pg{8116--8118}.

\bibitem[Garrick(1936)]{garrick1936propulsion}
{\sc \au{Garrick, I.~E.}} \yr{1936}  \at{Propulsion of a flapping and
  oscillating airfoil.}  \jt{NACA Tech. Rep.}  \bvol{567},  \pg{419--427}.

\bibitem[Green {\em et~al.\/}(2011)Green, Rowley \& Smits]{green2011unsteady}
{\sc \au{Green, M.~A.}, \au{Rowley, C.~W.} \& \au{Smits, A.~J.}} \yr{2011}
  \at{The unsteady three-dimensional wake produced by a trapezoidal pitching
  panel}.  \jt{Journal of Fluid Mechanics}  \bvol{685},  \pg{117--145}.

\bibitem[Kang {\em et~al.\/}(2011)Kang, Aono, Cesnik \& Shyy]{kang2011effects}
{\sc \au{Kang, C.~K.}, \au{Aono, H.}, \au{Cesnik, C. E.~S.} \& \au{Shyy, W.}}
  \yr{2011}  \at{Effects of flexibility on the aerodynamic performance of
  flapping wings}.  \jt{Journal of fluid mechanics}  \bvol{689},  \pg{32--74}.

\bibitem[King {\em et~al.\/}(2018)King, Kumar \& Green]{king2018}
{\sc \au{King, J.~T.}, \au{Kumar, R.} \& \au{Green, M.~A.}} \yr{2018}
  \at{Experimental observations of the three-dimensional wake structures and
  dynamics generated by a rigid, bioinspired pitching panel}.  \jt{Physical
  Review Fluids}  \bvol{3}~(034701).

\bibitem[Krasny(1986)]{krasny1986desingularization}
{\sc \au{Krasny, R.}} \yr{1986}  \at{Desingularization of periodic vortex sheet
  roll-up}.  \jt{Journal of Computational Physics}  \bvol{65}~(2),
  \pg{292--313}.

\bibitem[Mackowski \& Williamson(2015)]{mackowski2015direct}
{\sc \au{Mackowski, A.~W.} \& \au{Williamson, C. H.~K.}} \yr{2015}  \at{Direct
  measurement of thrust and efficiency of an airfoil undergoing pure pitching}.
   \jt{Journal of Fluid Mechanics}  \bvol{765},  \pg{524--543}.

\bibitem[Masoud \& Alexeev(2010)]{masoud2010resonance}
{\sc \au{Masoud, H.} \& \au{Alexeev, A.}} \yr{2010}  \at{Resonance of flexible
  flapping wings at low reynolds number}.  \jt{Physical Review E}
  \bvol{81}~(5),  \pg{056304}.

\bibitem[McCune \& Tavares(1993)]{mccune1993perspective}
{\sc \au{McCune, J.~E.} \& \au{Tavares, T.~S.}} \yr{1993}  \at{Perspective:
  unsteady wing theory—the k{\'a}rm{\'a}n/sears legacy.}  \jt{Journal of
  fluids engineering}  \bvol{115}~(4),  \pg{548--560}.

\bibitem[Mivehchi {\em et~al.\/}(2016)Mivehchi, Dahl \&
  Licht]{mivehchi2016heaving}
{\sc \au{Mivehchi, A.}, \au{Dahl, J.} \& \au{Licht, S.}} \yr{2016}  \at{Heaving
  and pitching oscillating foil propulsion in ground effect}.  \jt{Journal of
  Fluids and Structures}  \bvol{63},  \pg{174--187}.

\bibitem[Moored(2018)]{moored2018unsteady}
{\sc \au{Moored, K.~W.}} \yr{2018}  \at{Unsteady three-dimensional boundary
  element method for self-propelled bio-inspired locomotion}.  \jt{Computers \&
  Fluids}  \bvol{167},  \pg{324--340}.

\bibitem[Moored {\em et~al.\/}(2014)Moored, Dewey, Boschitsch, Smits \&
  Haj-Hariri]{moored2014linear}
{\sc \au{Moored, K.~W.}, \au{Dewey, P.~A.}, \au{Boschitsch, B.~M.}, \au{Smits,
  A.~J.} \& \au{Haj-Hariri, H.}} \yr{2014}  \at{Linear instability mechanisms
  leading to optimally efficient locomotion with flexible propulsors}.
  \jt{Physics of Fluids}  \bvol{26}~(4),  \pg{041905}.

\bibitem[Moored {\em et~al.\/}(2012)Moored, Dewey, Smits \&
  Haj-Hariri]{moored2012hydrodynamic}
{\sc \au{Moored, K.~W.}, \au{Dewey, P.~A.}, \au{Smits, A.~J.} \&
  \au{Haj-Hariri, H.}} \yr{2012}  \at{Hydrodynamic wake resonance as an
  underlying principle of efficient unsteady propulsion}.  \jt{Journal of Fluid
  Mechanics}  \bvol{708},  \pg{329--348}.

\bibitem[Moored \& Quinn(2018)]{moored2018inviscid}
{\sc \au{Moored, K.~W.} \& \au{Quinn, D.~B.}} \yr{2018}  \at{Inviscid scaling
  laws of a self-propelled pitching airfoil}.  \jt{AIAA Journal}  \pg{pp.
  1--15}.

\bibitem[Perkins {\em et~al.\/}(2017)Perkins, Elles, Badlissi, Mivehchi, Dahl
  \& Licht]{perkins2017rolling}
{\sc \au{Perkins, Matthew}, \au{Elles, Dane}, \au{Badlissi, George},
  \au{Mivehchi, Amin}, \au{Dahl, Jason} \& \au{Licht, Stephen}} \yr{2017}
  \at{Rolling and pitching oscillating foil propulsion in ground effect}.
  \jt{Bioinspiration \& biomimetics}  \bvol{13}~(1),  \pg{016003}.

\bibitem[Prandtl(1920)]{prandtl1920theory}
{\sc \au{Prandtl, L.}} \yr{1920}  \at{Theory of lifting surfaces.}  \jt{NACA
  Tech. Rep.}  \bvol{9},  \pg{1--11}.

\bibitem[Quinn {\em et~al.\/}(2014{\natexlab{{\em a\/}}})Quinn, Lauder \&
  Smits]{quinn2014scaling}
{\sc \au{Quinn, D.~B.}, \au{Lauder, G.~V.} \& \au{Smits, A.~J.}}
  \yr{2014{\natexlab{{\em a\/}}}}  \at{Scaling the propulsive performance of
  heaving flexible panels}.  \jt{Journal of fluid mechanics}  \bvol{738},
  \pg{250--267}.

\bibitem[Quinn {\em et~al.\/}(2014{\natexlab{{\em b\/}}})Quinn, Moored, Dewey
  \& Smits]{quinn2014unsteady}
{\sc \au{Quinn, D.~B}, \au{Moored, K.~W.}, \au{Dewey, P.~A.} \& \au{Smits,
  A.~J.}} \yr{2014{\natexlab{{\em b\/}}}}  \at{Unsteady propulsion near a solid
  boundary}.  \jt{Journal of Fluid Mechanics}  \bvol{742},  \pg{152--170}.

\bibitem[Saadat {\em et~al.\/}(2017)Saadat, Fish, Domel, Di~Santo, Lauder \&
  Haj-Hariri]{saadat2017}
{\sc \au{Saadat, M.}, \au{Fish, F.~E.}, \au{Domel, A.~G.}, \au{Di~Santo, V.},
  \au{Lauder, G.~V.} \& \au{Haj-Hariri, H.}} \yr{2017}  \at{On the rules for
  aquatic locomotion}.  \jt{Phys. Rev. Fluids}  \bvol{2},  \pg{083102}.

\bibitem[Sambilay~Jr(1990)]{sambilay1990interrelationships}
{\sc \au{Sambilay~Jr, V.~C.}} \yr{1990}  \at{Interrelationships between
  swimming speed, caudal fin aspect ratio and body length of fishes}.
  \jt{Fishbyte}  \bvol{8}~(3),  \pg{16--20}.

\bibitem[Senturk \& Smits(2019)]{Senturk2019}
{\sc \au{Senturk, Utku} \& \au{Smits, Alexander~J.}} \yr{2019}  \at{{Reynolds
  number scaling of the propulsive performance of a pitching airfoil}}.
  \jt{AIAA Journal}  \bvol{57}~(7),  \pg{2663--2669}.

\bibitem[Theodorsen(1935)]{theodorsen1935general}
{\sc \au{Theodorsen, T.}} \yr{1935}  \at{General theory of aerodynamic
  instability and the mechanism of flutter.}  \jt{NACA Tech. Rep.}  \bvol{496},
   \pg{413--433}.

\bibitem[Van~Buren {\em et~al.\/}(2018)Van~Buren, Floryan \&
  Smits]{van2018scaling}
{\sc \au{Van~Buren, T.}, \au{Floryan, D.} \& \au{Smits, A.~J.}} \yr{2018}
  \at{Scaling and performance of simultaneously heaving and pitching foils}.
  \jt{AIAA Journal}  \pg{pp. 1--12}.

\end{thebibliography}

\end{document}